\documentclass[%
reprint,
superscriptaddress,
amsmath,amssymb,
aps,
prb,
floatfix
]{revtex4-2}

\usepackage{multirow}
\usepackage{graphicx}
\usepackage{dcolumn}
\usepackage{bm}
\usepackage[version=3]{mhchem}
\usepackage{nicefrac}

\usepackage{color}
\usepackage{dcolumn}
\usepackage{amssymb}
\usepackage{url}
\usepackage{hyperref}
\usepackage{xfrac}
\usepackage{tabularx}
\usepackage{xspace}
\usepackage{amsmath}
\usepackage[normalem]{ulem}
\usepackage{mathtools}
\usepackage{cleveref}

\def\be{\begin{equation}}
\def\ee{\end{equation}}
\def\bea{\begin{eqnarray}}
\def\eea{\end{eqnarray}}
\def\ba{\begin{array}}
\def\ea{\end{array}}

\begin{document}
\title{Phase diagram of the $J$-$J_d$ Heisenberg Model on the Maple-Leaf Lattice: \\ Neural networks and density matrix renormalization group}
\author{Jonas Beck}
\affiliation{Institut f\"ur Theoretische Physik und Astrophysik and W\"urzburg-Dresden Cluster of Excellence ct.qmat, Universit\"at W\"urzburg,
Am Hubland Campus S\"ud, W\"urzburg 97074, Germany}
\author{Jonathan Bodky}
\affiliation{Institut f\"ur Theoretische Physik und Astrophysik and W\"urzburg-Dresden Cluster of Excellence ct.qmat, Universit\"at W\"urzburg,
Am Hubland Campus S\"ud, W\"urzburg 97074, Germany}
\author{Johannes Motruk}
\affiliation{Department of Theoretical Physics, University of Geneva, Quai Ernest-Ansermet 24, 1211 Geneva, Switzerland}
\author{Tobias M\"uller}
\affiliation{Institut f\"ur Theoretische Physik und Astrophysik and W\"urzburg-Dresden Cluster of Excellence ct.qmat, Universit\"at W\"urzburg,
Am Hubland Campus S\"ud, W\"urzburg 97074, Germany}
\author{Ronny Thomale}
\affiliation{Institut f\"ur Theoretische Physik und Astrophysik and W\"urzburg-Dresden Cluster of Excellence ct.qmat, Universit\"at W\"urzburg,
Am Hubland Campus S\"ud, W\"urzburg 97074, Germany}
\author{Pratyay Ghosh}
\email{pratyay.ghosh@epfl.ch}
\affiliation{Institut f\"ur Theoretische Physik und Astrophysik and W\"urzburg-Dresden Cluster of Excellence ct.qmat, Universit\"at W\"urzburg,
Am Hubland Campus S\"ud, W\"urzburg 97074, Germany}
\affiliation{Institute of Physics, Ecole Polytechnique F\'ed\'erale de Lausanne (EPFL), CH-1015 Lausanne, Switzerland}

\begin{abstract}
We microscopically analyze the nearest neighbor Heisenberg model on the maple-leaf lattice through neural quantum states (NQS) and infinite density matrix renormalization group (iDMRG). Embarking to parameter regimes beyond the exact dimer singlet ground state with a dimer bond spin exchange coupling $J_d$ varied against the exchange strength $J$ of all other bonds, iDMRG (NQS) finds a dimer state paramagnetic phase for $J_d/J > 1.464$ ($J_d/J > 1.39$) and a canted $120^\circ$ magnetic order for $J_d/J < 1.419$ ($J_d/J < 1.23$). Assessing training convergence inaccuracies of NQS and the influence of finite cylindric circumference for iDMRG, we discuss the possible existence of an intermediate phase between magnet and dimer paramagnet.
\end{abstract}\maketitle


\section{Introduction} The analysis of spin Hamiltonians suspected to yield phenomena of competing phases in frustrated magnetism is a notoriously difficult problem in the field of strongly correlated electron systems~\cite{doi:10.1146/annurev.ms.24.080194.002321,balents,Diepbook,frustrationbook}. Since the energy of a quantum spin state is dominated by local spin correlations while many ground states are similar in their short-range yet different in their long-range correlation profile, there are typically several competing candidate states. This makes it difficult not to fall for some kind of bias implied by mean field decoupling, effective models, or quasiparticle representation. As a consequence, finding the phase diagram of a frustrated quantum magnet Hamiltonian is often constrained to numerical microscopic approaches, where the calculation of energy densities and ground state correlation functions allows one to obtain some grip on the task. Ideally, it is desirable not to be limited to the exact diagonalization of finite size clusters either, whose system length might undergo the characteristic lengths of the unfolding ground state nature in the thermodynamic limit. 

An additional desirable feature for analyzing a magnetic quantum phase diagram is the existence of exactly known ground states at certain points or domains within the chosen parameter space, which then serve as a pivot to embark on regimes that are not exactly known. Under certain circumstances, systematic higher-order perturbative approaches around such a pivot point are already sufficient to detect phase transitions into adjacent phases~\cite{knetteruhrig,PhysRevB.80.081104}. Even from an all-numerical outset, the exact reference points or domains are valuable in order to benchmark a given method's performance, and provide further substantiation to the overall numerical results.

Analyzing the Heisenberg model on the maple-leaf lattice~\cite{Betts1995}, termed as maple-leaf model (MLM), is currently evolving into a vibrant subbranch of quantum magnetism. More than 40 years after the groundbreaking foundation of the Shastry-Sutherland model (SSM) featuring an exact dimer ground state~\cite{Shastry1981}, three of us have recently shown that the MLM features yet another exact dimer singlet ground state and that MLM and SSM are the only two instances for all lattices in two spatial dimensions with uniform tilings~\cite{ghosh2022another,Ghosh2023}. While the superlattice of dimer hopping dynamics forms a square lattice for the SSM, it forms a kagome superlattice for the MLM. Keeping in mind the significant interest the SSM phase diagram sparked over the past decades as to study the competition of magnetism and dimer paramagnets on most substantiated microscopic footing~\cite{Corboz2013} which culminated in the proposal of an intriguing magnetization plateau profile~\cite{PhysRevLett.112.147203} and, most recently, an exotic spin liquid phase in the SSM~\cite{Shi2022,Yang2022}, it naturally suggests the question which phase diagram is born out of such similar competition for the MLM. 

In this article, we apply two microscopic numerical approaches in order to retrieve information about the MLM phase diagram. First, we use group equivariant convolutional neural network algorithms applied to neural quantum states (NQS) to obtain ground state energies and spin correlations through an ansatz inspired by machine learning. Second, we employ infinite density matrix renormalization group (iDMRG) applied to infinite-length cylinders formed by maple-leaf unit cells with a finite circumference. While iDMRG is methodologically more established already, the NQS ansatz is witnessing increasing popularity and promises to benefit significantly from the across-the-board scientific excitement about the utilization of machine learning for scientific problem tasks. Both approaches allow us to calculate energy densities and static spin-spin correlators and as such guarantee a complete comparability within our analysis. The article is organized as follows. We introduce the MLM in Section~\ref{sec-model}, followed by our NQS and iDMRG approach in Section~\ref{sec-methods}. The MLM phase diagram as revealed in Section~\ref{sec-results} through NQS and iDMRG features a dimerized phase for dominant dimer bond coupling and canted $120^\circ$ magnetic order for subdominant dimer bond coupling. The key subtlety, where the approaches also differ the most, reveals itself for the intermediate regime between the two limits, the interpretation of which is particularly elaborated on in Section~\ref{sec-discussion}. In Section~\ref{sec-conclusion}, we conclude that while an intermediate phase interpolating between magnetism and dimer paramagnetism cannot be excluded, there is ambiguous evidence from NQS and iDMRG with regard to its nature and principle existence. 

\section{Model}\label{sec-model}
\begin{figure}[h] 
\begin{center}
\includegraphics[width=0.8\linewidth]{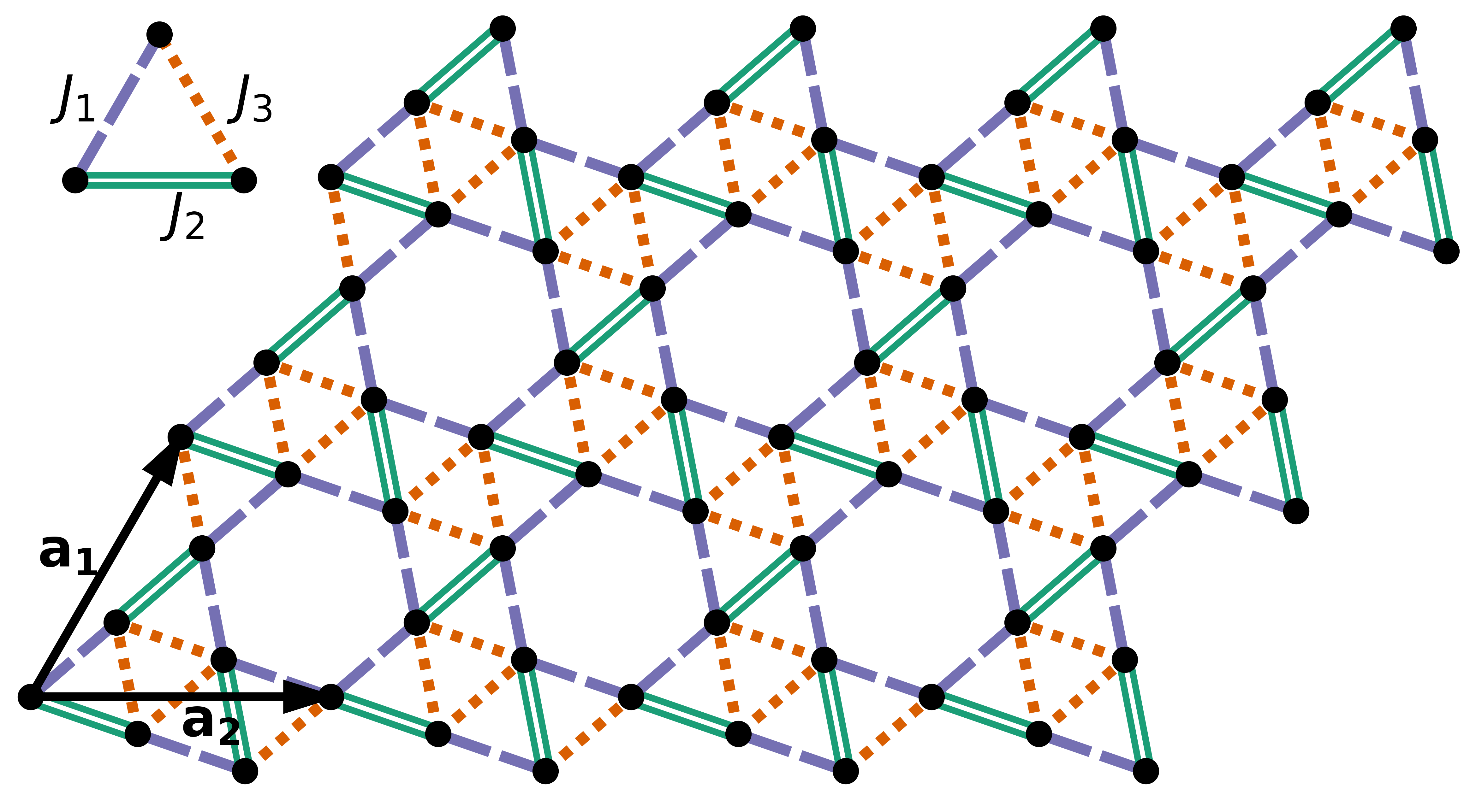}
\end{center}
\caption[]{Maple-leaf lattice with three symmetry-inequivalent nearest neighbour couplings $J_1$, $J_2$, and $J_3$. The green double-lines depict dimer couplings ($J_d\equiv J_2$) while violet (dashed) or red (dotted) bonds represent the inter-dimer couplings ($J\equiv J_1 \equiv J_3$), leading to the $J$-$J_d$ MLM.
}\label{fig:mllattice}
\end{figure}

The maple-leaf lattice (MLL)~\cite{Betts1995} has a uniform snub trihexagonal tiling, where each vertex is surrounded by four triangles and one hexagon (see Fig.~\ref{fig:mllattice}). The lattice is obtained by $1/7$-th site depletion of the regular triangular lattice~\citep{Betts1995} with a coordination number $z=5$. The lattice corresponds to a $p6$ plane group symmetry, referring to the sixfold rotational symmetry around the centers of the hexagons. It has three symmetry-inequivalent nearest neighbor bonds, which are marked in different colors and styles in Fig.~\ref{fig:mllattice}. The model we discuss in this article is the nearest-neighbor antiferromagnetic Heisenberg model on the maple-leaf lattice (MLM), which reads
\begin{equation}\label{eq:ml-ham}
\hat{\mathcal{H}}_\text{MLM}=
J_{1}\sum_{\langle i,j\rangle_1}\mathbf{\hat{S}}_{i}\cdot\mathbf{\hat{S}}_{j}
+ J_{2}\sum_{\langle i,j\rangle_2}\mathbf{\hat{S}}_{i}\cdot\mathbf{\hat{S}}_{j}
+ J_{3}\sum_{\langle i,j\rangle_3}\mathbf{\hat{S}}_{i}\cdot\mathbf{\hat{S}}_{j},
\end{equation}
where $\langle\rangle_k$ is a summation over nearest neighbors connected by a bond of type $k$ with a coupling strength $J_k$ as shown in Fig.~\ref{fig:mllattice}. $\widehat{\bold{S}}_i$ denotes the operator acting on a spin-$1/2$ representation on site $i$. In this article, we constrain ourselves to a specific subspace of the MLM with $J \coloneqq J_1=J_3$, denoting the non-dimer cover bonds, and $J_d\coloneqq J_2$ denoting the dimer cover bonds, as it was introduced by three of us in Ref.~\cite{ghosh2022another}. There it was analytically demonstrated that the $J$-$J_d$ MLM hosts an exact dimer ground state for $J_d >2J$. In that case, the ground state is a product of dimer singlets on all the $J_d$ bonds. The model has, in principle, been preconceived through some earlier numerical investigations~\cite{Farnell2011,Schmalfuss2002,Jahromi2020}. A comprehensive understanding of the phase diagram as a function of $J_d/J$, however, as of yet has neither been extensively pursued nor achieved. For the sake of simplicity, we set $J=1$ throughout the remainder of this article, and only depict parameter sweeps as a function of $J_d$. 

\section{Methods} \label{sec-methods}
We study the phase diagram of the $J$-$J_d$ MLM via Group Equivariant Convolutional Neural Network neural quantum states (NQS)~\cite{cohen2016} and the infinite density matrix renormalization group (iDMRG)~\cite{White1992,McCulloch2008}. Since the degree of maturity and ubiquitous use of iDMRG in the condensed matter research community is higher than for NQS, we emphasize explicating the NQS approach as we briefly introduce both techniques. 
\subsection{Group Equivariant Convolutional Neural Networks}
Due to the universal approximation theorem~\cite{Hornik1989}, neural networks (NN) can in principle represent any smooth function with arbitrary accuracy. This led Carleo and Troyer~\cite{Carleo2017} to propose their use as an unbiased parametrization function for variational quantum states. Such NQS have repeatedly been successfully employed to investigate the ground state properties of different quantum many-body systems~\cite{Choo2019,viteritti2023transformer,chen2023, Szabo2020, Hibat-Allah2020, Nomura2021, Sharir2020, mezera2023neural, Astrakhantsev2021, Viteritti2023PRL, Roth2023}. Within the abundance of different NN architectures available, we focus on group equivariant convolutional neural networks (GCNNs) \cite{cohen2016}. They are a generalization of convolutional neural networks and completely equivariant with respect to a given discrete symmetry group $G$, which means that their output can easily be enforced to transform according to any irreducible representation of $G$. Since our goal is to describe the ground states (GS) of solid-state systems, which necessarily follow the lattice symmetry, they are well suited for this application \cite{Roth2021, Roth2023, Duric2024}.
GCNNs are feed-forward networks and consist of layers of the form
\begin{equation}\label{eq:cnn-eq}
{f^{l}}_m(g) = z^l\Big\{ \sum_{r=1}^{F(l-1)} \, [{f^{l-1}}_r * {K^{l}}_{m, r}](g) + {b^l}_m \Big\},
\end{equation}
where $g\in G$. In the $l$-th layer ${f^{l}}_m$ is the $m$-th feature map with a corresponding convolution kernel ${K^{l}}_{m, r}$, connecting to the $r$-th feature map of the previous layer, and a bias ${b^l}_m$. Together, these kernels and biases form the complex network parameters $\alpha$. ${z^l:\,\mathbb{C}\to\mathbb{C}}$ is the nonlinear activation function (we use the Scaled Exponential Linear Unit ~\cite{klambauer2017selfnormalizing} applied separately to the real and imaginary part of its input), and $F(l)$ is the number of feature maps in the $l$-th layer. The first layer, which is called the embedding layer, maps a computational basis state $f^0(\vec{y})$, $\vec{y}\in\mathbb{Z}^2$, to feature maps over the symmetry group $G$ via the convolution
\begin{equation}\label{eq:gcnn-conv1}
[f^0 * K](g) = \sum_{\vec{y} \in \mathbb{Z}^2}\, f^0(\vec{y})K(g^{-1}\vec{y}),
\end{equation}
whereas the following layers map $G\to G$ via
\begin{equation}\label{eq:gcnn-conv2}
[f * K](g) = \sum_{h \in G}\, f(h)K (g^{-1}h).
\end{equation}
Finally, the output of the network, i.e. the wave function amplitude, is calculated in the output layer
\begin{equation}\label{eq:final-layer}
\Psi_\alpha (f^0) = {\sum_{\substack{r=1
}}^{F(l_\text{out})}\sum_{g \in G}\,\chi_g^* \, e^{{f^{l_\text{out}}}_r(g)}}
\end{equation}
with the characters $\chi_g$ corresponding to the desired irreducible representation of $G$, and $f^0$ an input state of the computational basis. This enforces the transformation of the variational state under the given irreducible representation \cite{Heine1960}, while the exponentiation allows for an easier representation of wave function amplitudes spanning several orders of magnitude.

As we aim to perform our calculation on systems with linear dimension, $L=3,6,9$ ($L$ unit-cells along each lattice vector, i.e. $N=L\times L\times6$ spin systems), we need GCNNs with enough parameters to sufficiently capture the complexity of the ground-state up to the largest system size. For this, we take up three different architectures of GCNNs. The first one, deemed as GCNN1, features four convolution layers with six feature maps each and an additional one with two feature maps before the mandatory symmetrization output layer. This results in $61262$ variational parameters for the $L=9$ lattice, $27242$ parameters for $L=6$ and $6830$ parameters on $L=3$ lattices. For smaller lattices, smaller and less deep networks are sufficient to achieve stable results. Therefore, for $L=6$ we also test GCNN2 with three convolution layers with respectively 6, 4, and 2 feature maps per layer ($8220$ parameters), and GCNN3 which is similar to GCNN2, but with respectively 8, 6, and 4 feature maps per layer and thus $17298$ parameters. Please take note that, like every variational wavefunction approach, achieving convergence on bigger system sizes becomes increasingly harder, as the Hilbert space dimension grows exponentially with $L$ while the number of parameters only grows polynomially for a fixed layer architecture.

The NQS is optimized via stochastic reconfiguration (SR)~\cite{sorellaSR}. This is a variational Monte Carlo method that iteratively minimizes the energy while, in each iteration, fulfilling the constraint of small distances between the old and new state, measured by the Fubini–Study metric tensor~\cite{Stokes2020quantumnatural}. SR can alternatively be interpreted as an imaginary time evolution~\cite{Yuan2019theoryofvariational}, which implies that this optimization algorithm can not get stuck in local minima - given the variational ansatz is sufficiently expressive to adequately cover the Hilbert space around the ground state. The energy (and later any observable of interest) is sampled with Markov chain Monte Carlo (MCMC).

For training, we use schedules that increase the number of MCMC samples (per iteration) during training while simultaneously the learning rate decays. The different networks are trained with different hyperparameters. For GCNN1, we use $2000-16000$ samples with a learning rate that varies from $0.02$ to $0.005$. For GCNN2 and GCNN3, the training is done with $33000$ samples with learning rates down to $0.0001$. It turns out that the GCNN1 requires around $2000$ iterations for convergence, whereas the rest converges within about $600$. Note that the GCNN1 produces the most stable results but requires the highest computational cost.

\subsection{iDMRG}
We perform our DMRG simulations mainly on an infinite cylinder with a circumference of $L_2=3$ unit cells (with limited $L_2=4$ data in the Appendix) where the periodic direction around the cylinder winds along the $\bm{a_1}-\bm{a_2}$ lattice vector. This amounts to a cylinder circumference of $3\sqrt{7} \approx 8$ ($\approx 10.6$ for $L_2=4$) lattice spacings. In order to converge to the true ground state close to the transition into the exact dimer state, we have to initialize the simulations with the ground state of a nearby parameter point descending from the lower $J_d$ side. Without it, the algorithm tends to converge to the \emph{local} energy minimum of the very lowly entangled exact dimer state and hence tends to overestimate the extent of the dimer phase. If present, this artifactual feature can easily be diagnosed by the formation of an unphysical jump in the energy as a function of $J_d$. In our calculations, we keep a matrix product states (MPS) bond dimension of $\chi=1600$ up to $6400$ resulting in truncation errors below  $ 3 \cdot 10^{-5}$.

\begin{figure}
\begin{center}
\includegraphics[width=\columnwidth]{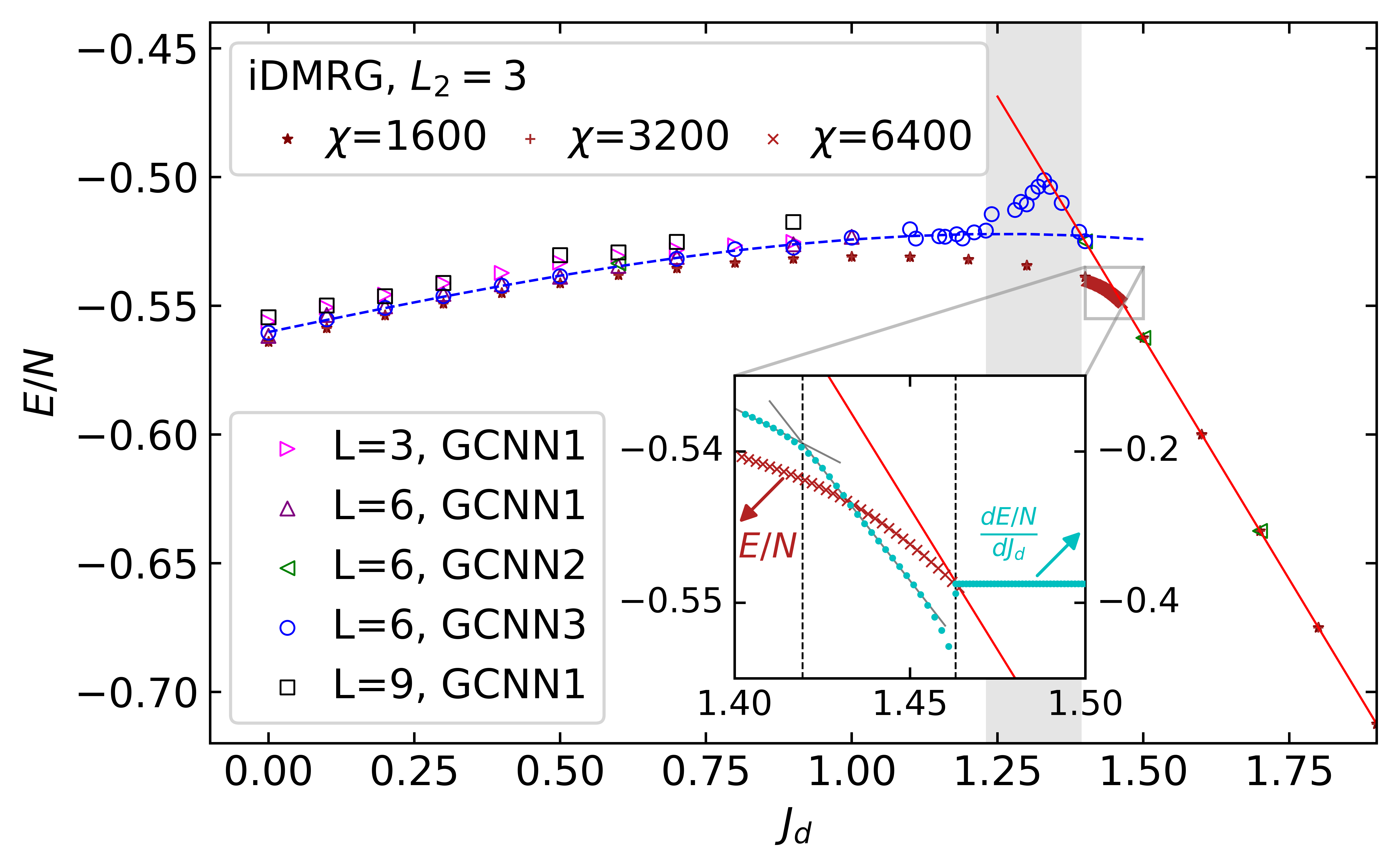}
\end{center}
\caption[]{GCNN and iDMRG results for the ground-state energy per site, $E/N$, as a function of $J_d$. $L$ is the number of unit cells in one direction, i.e. $L=3$ ($N=54$ spins), $L=6$ ($N=216$ spins), and $L=9$ ($N=486$ spins) for GCNN. GCNN1 is a five-layer network with 6, 6, 6, 6, and 2 feature maps per layer, GCNN2 is a three-layer network with respectively 6, 4, and 2 feature maps and GCNN3 is another three-layer network with 8, 6, and 4 feature maps. We plot iDMRG data with $L_2=3$ unit cells around the cylinder for comparison. In the grey area, the GCNNs show bad convergence and are significantly outperformed by the iDMRG calculations. The solid red line marks the energy of the exact dimer state, and the dashed blue line is an extrapolation of the GCNN3 data to get an estimation of the critical point at $J_\text{d,crit} = 1.39$. The Inset shows the iDMRG data at the phase transitions, suggesting an intermediate state for $1.419 < J_\text{d} < 1.464$.}
\label{fig:phase_diagram}
\end{figure}

\section{Results} \label{sec-results}

\subsection{Dimerized phase} 
The model in~\eqref{eq:ml-ham} possesses an exact dimer singlet ground state for $J_d>2$ \cite{ghosh2022another}, which is faithfully reproduced by NQS and iDMRG. The analytical calculations are based on the variational principle and thus do not exhaust the actual range of product singlet phase for $J_d<2$. Our GCNN calculations find the singlet dimer ground state for $J_d\ge 1.33(1)$ while iDMRG obtains $J_d \ge 1.464(2)$. Due to the locality of entanglement inherent to a dimer state, it is likely that the DMRG performs excellently in such a domain. The discrepancy between NQS and iDMRG rather appears to be stemming from the NQS not converging to the correct ground state near the critical regime.  Above $J_d=1.33$, the GCNNs, enforcing the trivial irreducible representation, consistently converge to the exact dimer eigenstate, while below $J_d=1.33$ the trivial irreducible representation GCNNs randomly choose between two energy bands to converge to, both of which are likely not the ground state of the system (Fig.~\ref{fig:phase_diagram} only depicts the lowest energy achieved for each $J_d$). In both scenarios, the non-trivial irreducible representations only converge when pre-trained with the trivial one, but never achieve better energies. As a result, ascending from the small $J_d$ side to the critical regime, the energy shows an inflection at $J_d=1.23(1)$, which meets with the singlet dimer phase at a cusp at $J=1.33(1)$ (Fig.~\ref{fig:phase_diagram}). In this region, for NQS, we choose to perform our calculations for $L=6$. This is because within our training protocols, all GCNNs perform best on the $L=6$ lattice ($216$ spins), i.e. it is most stable with the least amount of samples. From extrapolating the curve below the inflection point we can estimate the curve to intersect the dimer energy curve at $J_d=1.39(1)$, which can be interpreted as the critical point out of the exact dimer phase as expected from NQS. 

\subsection{Magnetic order} 
NQS and iDMRG reach unanimous evidence for magnetic order for $J_d<1.23$.
\begin{figure}
\includegraphics[width=0.8\columnwidth]{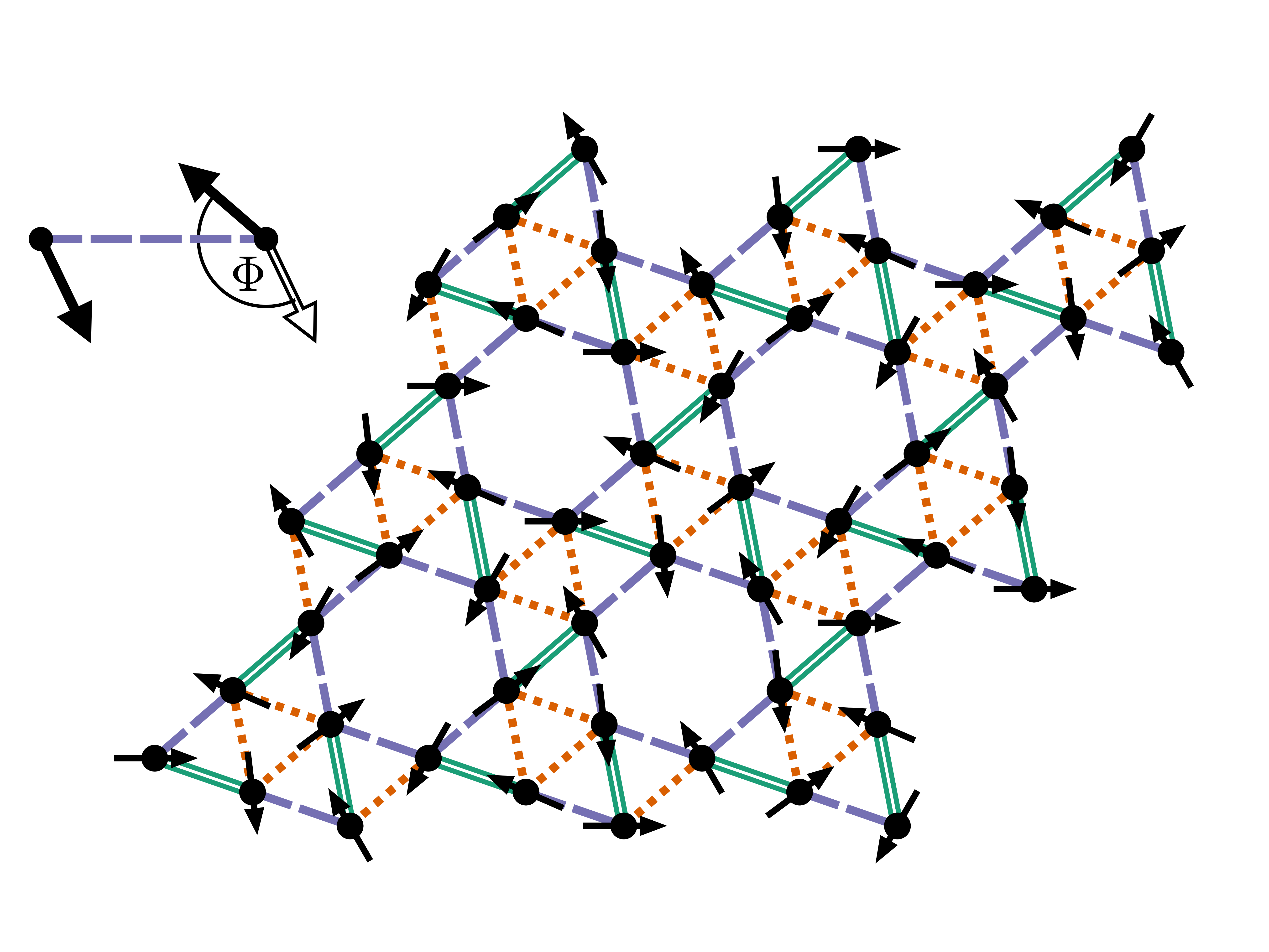}
\caption[]{Spin orientations in the classical limit at $J_d=0.8$. In this canted 120° order \cite{ghosh2022another}, spins on the red triangles show a 120° order, while spins across violet bonds are canted by a $J_d$-dependent canting angle $\Phi$. }
\label{fig:c120}
\end{figure}
To identify the ground state(s) within NQS, we calculate the sublattice magnetization given by
\begin{equation}
m(L) = \left[\frac{1}{6N_{uc}^2}\,\sum_{l,m=1}^{N_{uc}}\,\sum_{k=1}^{6}\,\langle\mathbf{\hat{S}}_{lk}\cdot \mathbf{\hat{S}}_{mk}\rangle \, e^{i \mathbf{Q} \cdot (\mathbf{R}_{l}-\mathbf{R}_{m})} \right]^{\nicefrac{1}{2}}
\end{equation}
where $l$ and $m$ sums run over all $N_{uc}$ unitcells and $k$ runs over the sublattices. $\mathbf{R}_{i}$ is the position of the $i$-th unitcell. Here, the ordering wavevector, $\mathbf{Q}=(\frac{8\pi}{3\sqrt{7}} , \frac{4\pi}{\sqrt{21}})$, corresponds to the classical canted-$120^\circ$ (c-$120^\circ$) order presented in Refs.~\cite{ghosh2022another,Farnell2011,Schmalfuss2002} and in Fig.~\ref{fig:c120}. The classical c-$120^\circ$ can be viewed as individual local $120^\circ$ order on the $J_3$ triangles with a relative canting of the spins between two neighboring triangles. The results for $J_d=0$, $0.3$, $0.7$, and $0.9$ for all three different system sizes, namely $L=3$, $6$, and $9$, are shown in Fig.~\ref{fig:magnetization}. The sublattice magnetizations for a fixed $J_d$ are extrapolated using $m(L)=c_0+c_1(1/L)^2$ to obtain an estimate of $m(L\to\infty)$. It is found that the $m(L\to\infty)$ is finite for the values $J_d$ mentioned above. This, along with the absence of any signature of a phase transition in the energy, indicates the classical c-$120^\circ$ to prevail in the region for $J_d<1.23$. In Fig.~\ref{fig:magnetization}, we also overlay NQS data of $m$ with that obtained from our iDMRG calculations, showing that both methods match approximately. The magnetization from iDMRG shows an upturn when approaching the critical regime. Apparently, there is no intuitive physical reason for that behavior and appears to be attributable to a finite-size artifact, cf.~Fig.~5 in~\cite{Farnell2011}.

\begin{figure}
\begin{center}
\includegraphics[width=\columnwidth]{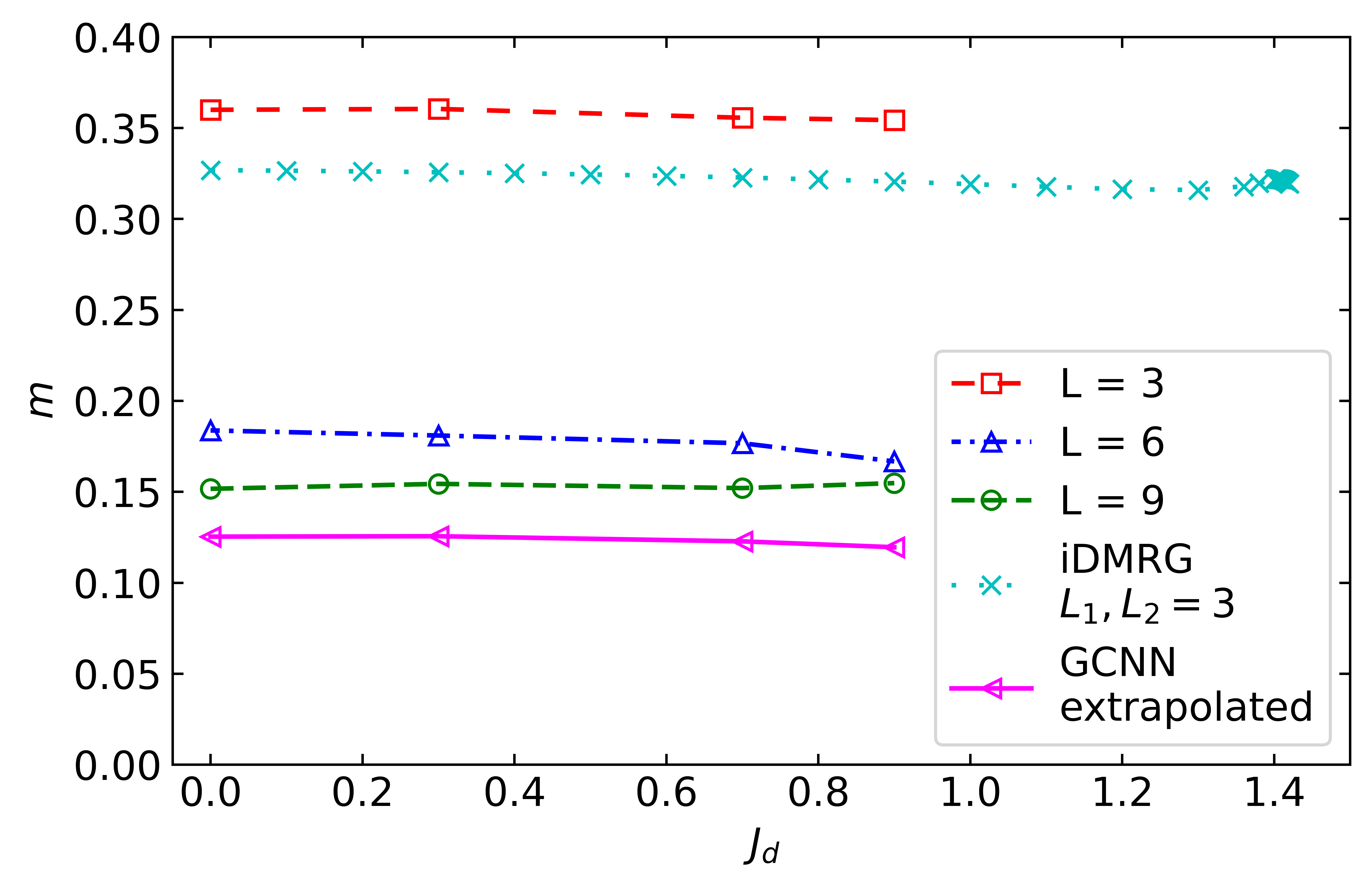}
\end{center}
\caption[]{GCNN results for the magnetic order parameter (sublattice magnetization) $m$ as a function of $J_d$. The extrapolated values in the limit $L \to \infty$ are calculated with a quadratic fit of the form $m(L) = c_0 + c_1 (1/L)^2$. For comparison, we also plot the iDMRG data within the ordered phase, calculated on a cluster of linear extent $L_1=3$ of the infinite cylinder of circumference $L_2=3$ unit cells.}
\label{fig:magnetization}
\end{figure}

\begin{figure*}
\includegraphics[width=\textwidth]{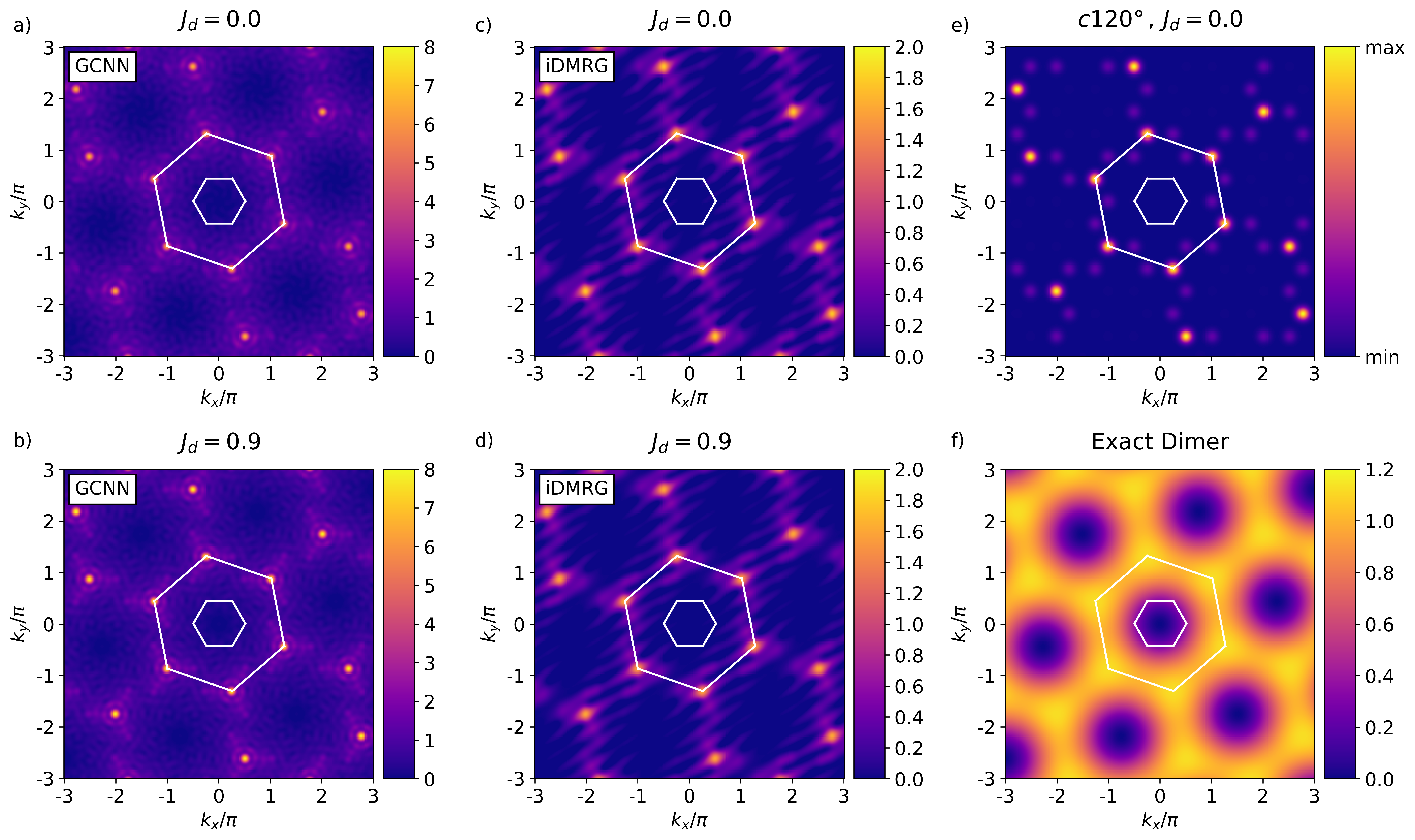}
\caption[]{Structure factors within the ordered phase. a) and b) show data from the $L=9$ GCNN1, c) and d) show iDMRG data with $L_2=3$ unit cells around the cylinder, e) is the classical c120 order at $J_d=0.0.$ simulated for $L=18$ and blurred with a gauss filter for better visibility of the peaks, and f) is the structure factor of the exact dimer state, both from \cite{ghosh2022another}.}
\label{fig:sf0}
\end{figure*}

The presence of magnetic order is also confirmed by the sharp Bragg peak-like features in the static structure factor, 
$$S(\mathbf{k})= \frac{1}{N}\,\sum_{l,m=1}^{N}\langle\mathbf{\hat{S}}_{l}\cdot \mathbf{\hat{S}}_{m}\rangle e^{i\mathbf{k}\cdot(\mathbf{r}_{l}-\mathbf{r}_{m})},$$
($\mathbf{r}_{i}$ is the position of the $i$-th site) shown in Fig.~\ref{fig:sf0}~(a) and (b) for NQS and in Fig.~\ref{fig:sf0}~(c) and (d) for iDMRG. The $S(\mathbf{k})$ found in numerical calculations are identical to the $S(\mathbf{k})$ in Fig.~\ref{fig:sf0}~(e) calculated for the c-$120^\circ$ order with classical spins. These peaks of $S(\mathbf{k})$ appear at $\mathbf{q}=(\nu_1\mathbf{a}_1^\star+\nu_2\mathbf{a}_2^\star)/3$, with $\nu_1$, $\nu_2$ being integers such that $\mod(\nu_1,3)\ne0$, $\mod(\nu_2,3)\ne0$, and $\mod(\nu_1+\nu_2,3)=\mod(2\nu_1+\nu_2,7)=0$, and $\mathbf{a}_1^\star$ and $\mathbf{a}_2^\star$ are the primitive vectors reciprocal to $\mathbf{a}_1$ and $\mathbf{a}_2$. The peaks coincide with the corners of the extended Brillouin zone determined by the vectors $\bm{\alpha}_1^\star$ and $\bm{\alpha}_2^\star$ which are the reciprocal vectors of $\bm{\alpha}_1=-\frac{1}{7}\mathbf{a}_1+\frac{3}{7}\mathbf{a}_2$ and $\bm{\alpha}_2=\frac{2}{7}\mathbf{a}_1+\frac{1}{7}\mathbf{a}_2$, where $\bm{\alpha}_1$ and $\bm{\alpha}_2$ are the lattice vectors of the underlying triangular lattice which is $1/7$-th site depleted to reach the maple-leaf lattice. The features of the c-$120^\circ$ order are \emph{almost} identical to the coplanar state on the triangular lattice~\cite{Messio-classical}. The difference is that the $S(\mathbf{k})$ in Fig.~\ref{fig:sf0}~(a) - (e), i.e. for both classical and quantum spins, features some satellite peaks (with lesser intensity) in addition to the main peaks. Each main peak is associated with three satellite peaks. For instance, the main peak at $\frac{5}{3}\mathbf{a}_1^\star+\frac{4}{3}\mathbf{a}_2^\star$ comes with three satellite peaks at $\mathbf{q}'=\frac{4}{3}\mathbf{a}_1^\star+\frac{2}{3}\mathbf{a}_2^\star$, $\frac{4}{3}\mathbf{a}_1^\star+\frac{5}{3}\mathbf{a}_2^\star$, and $\frac{7}{3}\mathbf{a}_1^\star+\frac{5}{3}\mathbf{a}_2^\star$. The relative intensity of these secondary peaks, i.e. $I_{\mathbf{q}'}/I_{\mathbf{q}}$, are directly related to the canting angle; $I_{\mathbf{q}'}/I_{\mathbf{q}}\approx 0.18$ for $J_d=0$ (canting angle $\Phi=\pi$), which monotonically decreases with increasing $J_d$ and approaches zero as $J_d\rightarrow\infty$ ($\Phi\rightarrow 2\pi/3$). We cannot faithfully obtain such ratios from NQS and iDMRG due to noise and finite-size artifacts, while we do expect that this particular ratio can be used to deduce the canting angle. In Fig.~\ref{fig:sf0}~(f), we also show the structure factor for the exact singlet dimer phase. The exact singlet dimer phase features a profile substantially different from the Bragg peaks, while soft maxima appearing on top of the broad background do occur at the same $\mathbf{q}$-vectors as the peaks in the $120^\circ$ order. 

\subsection{Intermediate regime} 
From NQS we could not determine the ground state for $1.23<J_d<1.39$. This is because as we investigated the model with GCNN3 on the $L=6$ lattice, we encountered difficulties during training as well as with convergence to the ground state (GS). For $J_d<1.23$ and $J_d>1.39$ the GCNNs agree with the iDMRG results, which led us to label the intermediate, as of now ambiguous, regime as $1.23<J_d<1.39$ (grey shaded domain in Fig.~\ref{fig:phase_diagram}). While the problematic performance of NQS in this regime is systematically evident from the training logs, there remains a generic ambiguity about the nature of the ground state near the critical point of the exact dimer phase, as not only NQS performs relatively poorly, but also iDMRG faces challenges: on the $L_2=3$ cylinder, iDMRG finds the c-$120^\circ$ from $J_d=0$ up to $J_d=1.42$, which is followed by a second order phase transition to a phase of unknown nature. This phase appears in the range $1.419<J_d<1.464$, as determined from a cusp in the derivative of the energy (inset of Fig.~\ref{fig:phase_diagram}), and we find that the correlation length shows a peak around this transition (see Appendix), just before the system undergoes a first order phase transition to the exact dimer phase at $J_d=1.464$. 
At the latter, we also find a jump in the entanglement entropy characteristic for a first-order transition, as one would generally expect for a phase transition involving an exact singlet phase~\cite{deconf-SSM,Corboz2013,Yang2022,ghosh2022another,Ghosh_QC}.

\section{Discussion}\label{sec-discussion}
It is difficult to provide a conclusive answer to the existence of an exotic phase sandwiched between the singlet dimer and the c-$120^\circ$ phase. With the help of coupled cluster approaches and exact diagonalization, Farnell \textit{et. al.}~\cite{Farnell2011} have suggested the absence of such a phase, whereas Gresista \textit{et. al.}~\cite{PhysRevB.108.L241116} argue in favor of it from pseudofermion functional renormalization group (PFFRG)~\cite{PhysRevB.81.144410,PhysRevB.83.024402,PhysRevB.89.020408,mueller2023pseudofermion}, and claim to find a quantum spin liquid in the $J$-$J_d$ MLM for $1.32<J_d<1.6$. Finite-size clusters might fall for their finite sizes when comparing competing ground states, while PFFRG does not obtain ground state energy densities, and only resorts to the static spin-spin correlator. Furthermore, PFFRG tends to overestimate paramagnetic regimes when the incipient magnetic ordering undergoes its resolution strength dependent on system size and frequency resolution. At our current stage of interpretation from our data, we find more preliminary indications in favor of a single critical point in line with Farnell \textit{et al.}, but cannot make any conclusive statement.

\begin{figure}
\includegraphics[width=0.8\columnwidth]{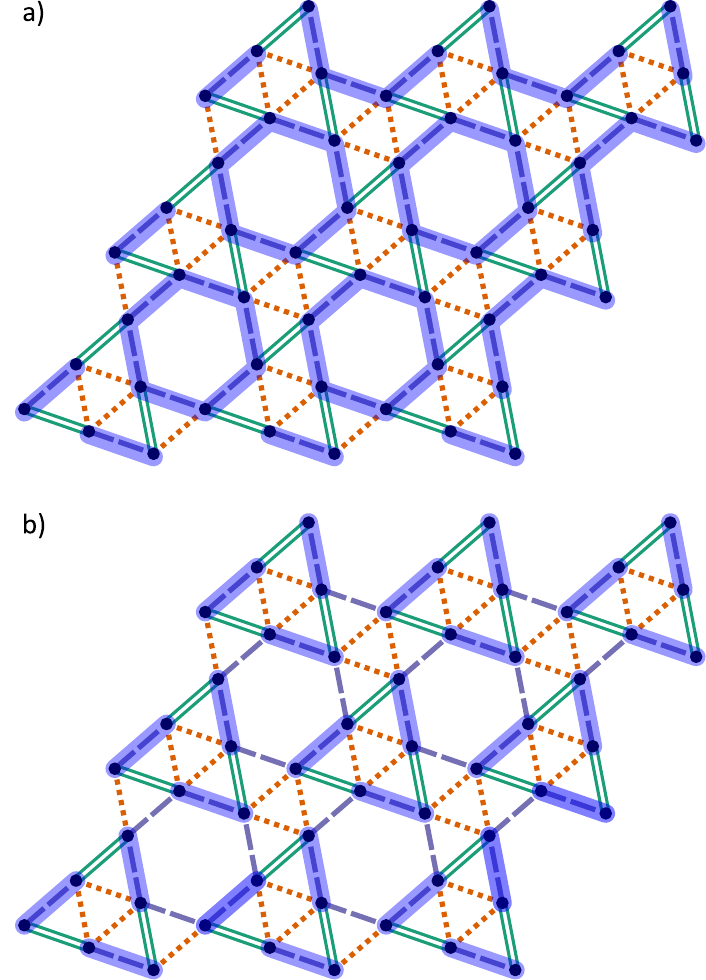}
\caption[]{The two possible candidate states that can appear between the exact singlet dimer phase and the magnetically ordered c-$120^\circ$ phase. The thick light blue hexagons/bonds depict strong singlet amplitude. (a) shows a plaquette VBS state. This state doe not break any lattice symmetries, whereas the dimer VBS shown in (b) breaks lattice rotation symmetry. The appearance of (b) can engender exotic criticality and/or a quantum spin liquid phase.}
\label{fig:candidate}
\end{figure}

To approach the hypothesis of an intermediate phase phenomenologically, however, it appears helpful to start from the MLM's similarity to the Shastry-Sutherland model (SSM)\cite{Shastry1981}. In the SSM, there exists a plaquette valence bond solid (VBS) phase between the classical N\'eel order and the exact dimer phase~\cite{Koga2000,Corboz2013,deconf-SSM}. This plaquette state in the Shastry-Sutherland model breaks lattice translation symmetry and shows a second order phase transition to the N\'eel ordered phase~\cite{Koga2000}. Such a phase transition can be associated with an exotic deconfined quantum criticality~\cite{deconf-SSM,Xi2023} or can hatch an intermediate quantum spin liquid phase~\cite{Yang2022,Liu2022,viteritti2023transformer}. As the MLM is more frustrated than the SSM, one can expect novel physics in the phase diagram. The analog for this SSM plaquette singlet state in the MLM would be the one depicted in Fig.~\ref{fig:candidate} (a); a state with strong singlet weight on the $J_1$ hexagons. Note that this state does not break any lattice symmetries, therefore a transition out of this phase to the magnetically ordered phase falls in the Ginzburg–Landau paradigm and does not necessarily conceive a quantum spin liquid. The other candidate state which might provide an exotic criticality or a quantum spin liquid is the dimer valence bond solid state shown in Fig.~\ref{fig:candidate} (b) which breaks the lattice rotation symmetry~\cite{ghosh2023effective}. Neither of our NQS and iDMRG calculations strongly indicate the appearance of such a state. The iDMRG spin-spin correlations for the VBS states obtained for $L_2=3$ and $4$ are shown in the Appendix. A faithful comparison appears challenging for the given finite cylinder circumference since the $L_2=4$ system is incommensurate with c-$120^\circ$ magnetic order, while all VBS states we find feature strong singlet amplitudes on all or some of the $J_d$ bonds. We thus cannot exclude the possibility that the intermediate phase in iDMRG is an artifact of the finite width of our system, which might disappear upon finite size scaling. From iDMRG, this would leave us with a first-order phase transition out of the exact dimer singlet phase directly to the c-$120^\circ$ magnetic order at $J_d\approx1.4$. NQS indicates this transition to occur at $J_d=1.39$. Such findings were in good agreement with the coupled cluster and exact diagonalization results obtained by Farnell et al.~\cite{Farnell2011}, which suggest a similar setting at $J_d\approx1.45$. 
The static spin structure factor ($S(\mathbf{k})$) of the MLM might further be interpreted as an indication of the absence of an intermediate phase. Note that the primary Bragg peaks of the c-$120^\circ$ and the soft maxima of the exact singlet phase are at the same points in the reciprocal space. In the classical ($S\to\infty$) SSM, the N\'eel state undergoes a phase transition to a spin spiral phase when the dimer interaction is increased~\cite{Shastry1981}. This spin spiral phase, upon the inclusion of quantum fluctuations, gives rise to the dimer singlet phase. Therefore, the $S(\mathbf{k})$ of the N\'eel and the exact singlet phases have no correspondence like the one we find for the MLM. There, the exact dimer singlet phase naturally evolves out of the c-$120^\circ$ order itself. 

An increase in methodological performance might also shed further light on the vexing questions related to the MLM phase diagram. In particular, on the NQS there appears to be significant room for improvement.
A major refinement of the GCNN can be accomplished by applying the MinSR~\cite{chen2023, rende2023} optimization algorithm, which reformulates SR to solve the major bottleneck of having to invert a large singular matrix. Thus, MinSR is significantly faster, more stable, and additionally allows for orders of magnitude more variational parameters for comparable computational cost. Another potential improvement to NQS is to use a visual transformer architecture, as recently proved successful for its application to the SSM~\cite{viteritti2023transformer}. These networks lack the advantage of exploiting symmetries but appear to make up for it with increased generality and sensitivity for correlations. Likewise, increasing the finite cylindric circumference appears to be a crucial bottleneck to overcome in order to increase the predictability of iDMRG to avoid unwanted incommensurability of a subset of candidate ground states. However, the bond dimensions needed for a reasonably accurate state description at the next larger favorable circumference of $L_2=6$ seem to be computationally out of reach. An alternative could be the application of inherently two-dimensional tensor networks~\cite{Corboz2013}, although the large unit cell and lattice geometry of the MLM might prove challenging.

\section{Conclusion}\label{sec-conclusion}
We have explored the phase diagram of the nearest neighbor antiferromagnetic Heisenberg model on the maple-leaf lattice (MLM) using neural quantum state (NQS) and infinite density matrix renormalization group (iDMRG) techniques. Our study is focused on the quantum phase diagram of the $J$-$J_d$ MLM~\cite{ghosh2022another,Ghosh2023}, for which we find canted $120^\circ$ magnetic order and a dimerized phase surrounding a potential intermediate phase or phase transition which we cannot conclusively resolve. While several lines of phenomenological reasoning might in fact hint at a somewhat simple MLM phase diagram without such an intermediate phase, the critical regime deserves further investigation. Our study provides valuable insights into the nature of quantum phases of the MLM, also showcasing the strengths and limitations of NQS and iDMRG in different regions of the phase diagram. The intriguing behavior near the critical domain underscores the complexity of quantum phase transitions on this non-trivial lattice, motivating future research to deepen our understanding of these phenomena. In particular, the simplification of the MLM to its parametric $J$-$J_d$ trajectory might hide the underlying complexity of the general MLM phase diagram.

\textit{Acknowledgments.}
We thank Fr\'ed\'eric Mila and Markus Heyl for useful discussions. This work is supported by the Deutsche Forschungsgemeinschaft (DFG, German Research Foundation) through Project-ID 258499086-SFB 1170 and the Würzburg-Dresden Cluster of Excellence on Complexity and Topology in Quantum Matter – ct.qmat Project-ID 390858490-EXC 2147.
J.M. acknowledges funding through the SNSF Swiss Postdoctoral Fellowship grant 210478. GCNN implementation and training was based on the \texttt{netket} library~\cite{netket3:2022, netket2:2019, mpi4jax:2021}. DMRG simulations were performed using the \texttt{TeNPy} library~\cite{tenpy} on the Boabab HPC cluster at the University of Geneva. The authors from W\"urzburg acknowledge the Gauss Centre for Supercomputing e.V. for providing computing time on the GCS Supercomputer SuperMUC at Leibniz Supercomputing Centre (LRZ).

\section*{Appendix}

\begin{figure*}[b!]
\includegraphics[width=\textwidth]{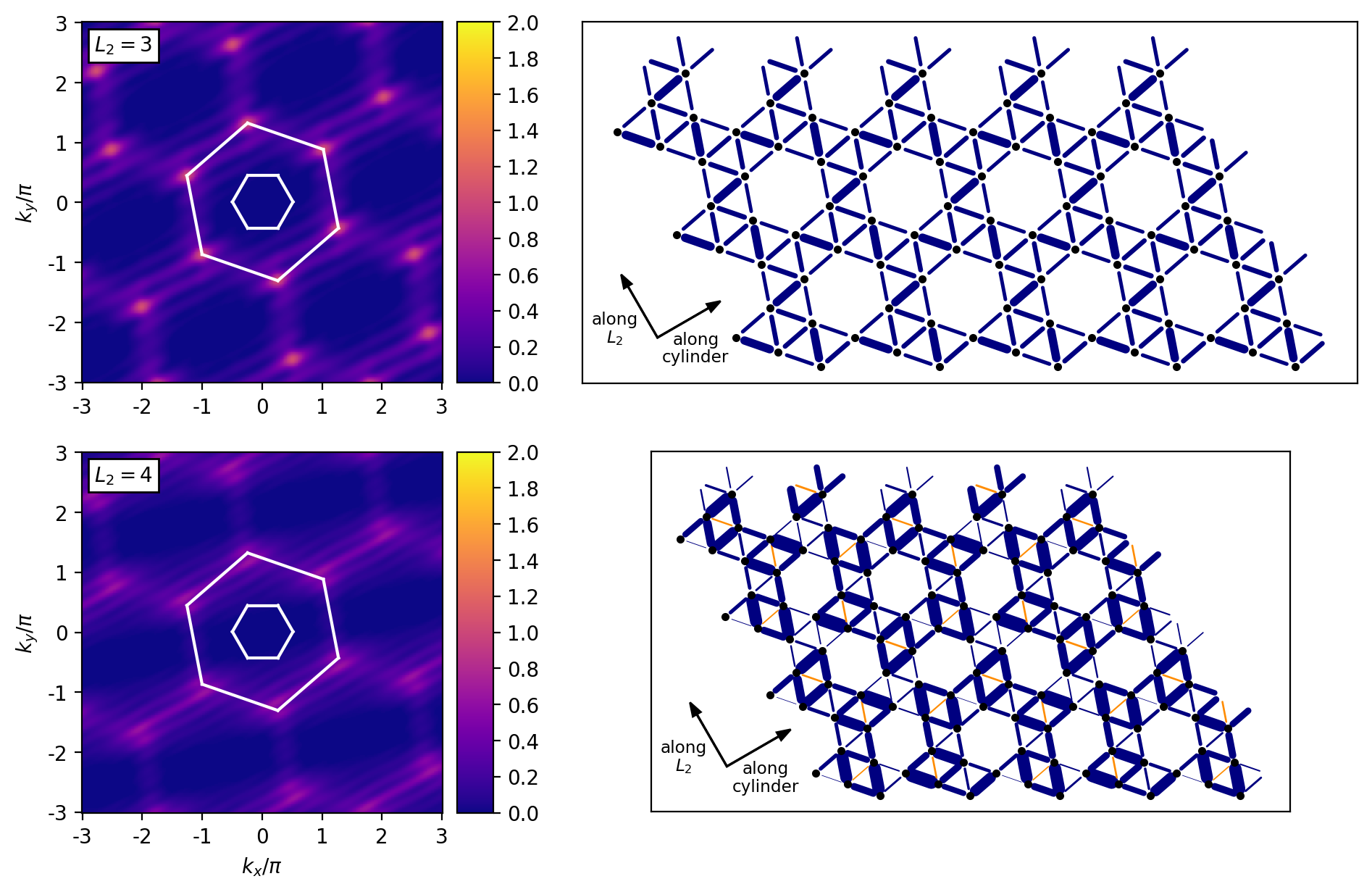}
\caption[]{Minimal energy states found by iDMRG close to the transition point into the exact dimer phase at $J_d=1.44$ for circumferences $L_2=3$ (upper panels) and $L_2=4$ (lower panels). The plots on the left show the structure factor. On the right, we depict the spin correlations across the nearest-neighbor bonds with the thickness of the lines being proportional to the respective correlation. Blue lines indicate negative (antiferromagnetic) correlations while orange lines imply positive (ferromagnetic) correlations.}
\label{fig:int_phase}
\end{figure*}

In Fig.~\ref{fig:int_phase}, we show the structure factors and spin correlations across nearest-neighbor bonds at $J_d=1.44$ we find with iDMRG for $L_2=3$ and $4$. 
This value of $J_d$ lies in the intermediate phase as determined on the $L_2=3$ geometry. The structure factor still exhibits maxima at the corners of the extended Brillouin zone, albeit lower and much less sharp than in the magnetically ordered c-120$^\circ$ phase. The bond correlations show vastly different behavior. For $L_2=3$, they are all antiferromagnetic and are largest on the $J_d$ bonds. 
In the $L_2=4$ system, on the other hand, sort of a pinwheel diamond valence bond crystal (VBC) pattern is forming with small ferromagnetic correlations on the bond inside the diamonds. As already mentioned in the in the main text, it is difficult to unambiguously determine the nature of this intermediate phase given the large unit cell of the maple-leaf lattice and the concomitant small number of accessible cylinder circumferences in iDMRG. The c-120$^\circ$ order is only commensurate for $L_2= 3n$ with $n$ integer meaning it is frustrated for $L_2=4$ while the appearing diamond pinwheel VBC is frustrated for $L_2\neq 2n$, hence including $3$. 

\begin{figure*}
\includegraphics[width=\textwidth]{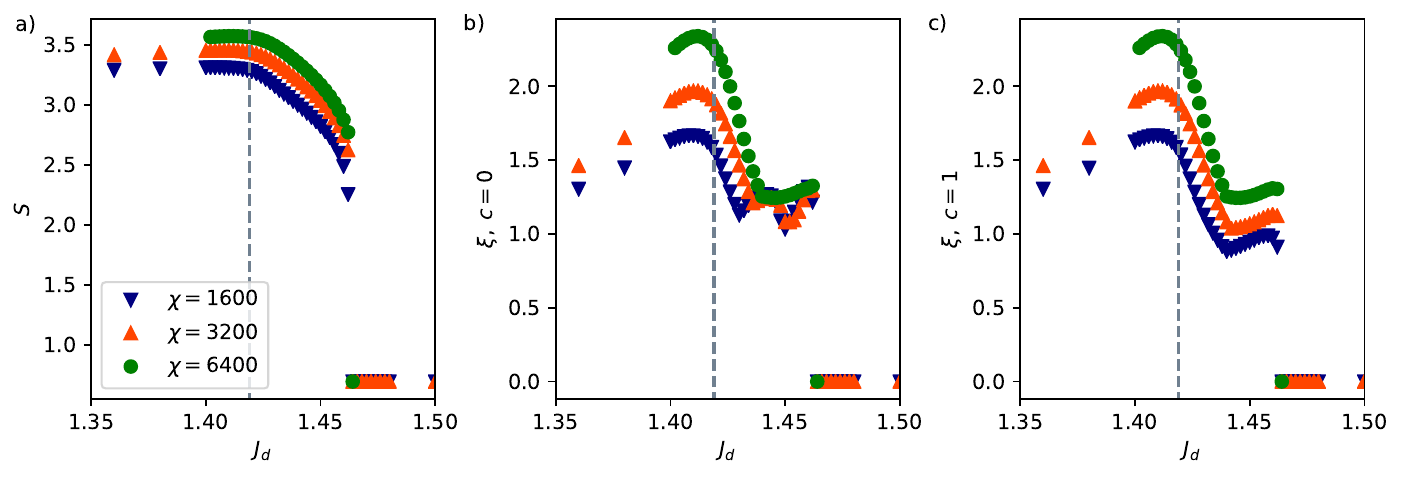}
\caption[]{a) entanglement entropy, b) correlation length of charge $0$, and c) correlation length of charge $1$ from iDMRG on the $L_2=3$ system. The gray dashed lines indicate the phase boundary determined from the derivative of the energy $E$ in Fig.~\ref{fig:phase_diagram}. The entanglement entropy decreases for higher values of $J_d$ and the correlation lengths show clear peaks in the vicinity of the phase boundary. Note that the peak is still moving to the right with increasing bond dimension $\chi$. A local quantity like $E$ will converge much faster with bond dimension compared to the highly nonlocal correlation length so that we consider the value from the energy derivative a more accurate estimation of the phase transition point on this geometry.}
\label{fig:iDMRG_S_xi}
\end{figure*}

In order to further corroborate the existence of an intermediate phase on the $L_2=3$ cylinder, we investigate entanglement entropy $S$ and correlations lengths $\xi$ along the cylinder depicted in Fig.~\ref{fig:iDMRG_S_xi}. $S$ in panel a) shows a clear downturn when exiting the c-120$^\circ$ phase at $J_d = 1.419$. The in two dimensions $SU(2)$ symmetry breaking c-120$^\circ$ phase is gapless with power law correlations and therefore has high entanglement entropy. In panel b) and c), we plot $\xi$ for charge $c=0$ and charge $c=1$, respectively. 
The charges here signify the change in charge when acting with some operator and the correlation functions of this operator fall of with at least this correlation length. Since the $U(1)$ charge in our simulations is $S^z$, $\xi, c=0$ e.g. governs the longest range correlations of $S^z_i S^z_j$ while $\xi, c=1$ determines the range of $S^+_i S^-_j$, with the latter being equivalent to the $x/y$ spin correlations. First of all, both $\xi$ show a clear peak growing with bond dimension, a signature for a continuous phase transition. The peak is still shifting towards higher $J_d$ with increasing bond dimension $\chi$ which makes it plausible that it would coincide with the transition point determined by the energy derivative in Fig.~\ref{fig:phase_diagram} (gray dashed line) in the infinite $\chi$ limit.  Second, the two $\xi$ have exactly the same value left of the transition consistent with the symmetric ground state of an $SU(2)$ symmetry breaking magnetically ordered phase in which the spin correlations in the $x$, $y$ and $z$ direction should be equivalent. In the intermediate phase,  $\xi, c=1$ is still growing with $\chi$ while $\xi, c=0$ seems to have converged, clearly indicating a different nature of the ground state. Around $J_d=1.46$, all three quantities show a clear jump indicating the first-order phase transition into the exact dimer state.

%

\end{document}